\documentclass[12pt,preprint]{aastex}
\usepackage{graphicx}

\usepackage{lscape}

\shorttitle{}

\shortauthors{Authors}

\begin{document}

\title{State-Dependent Orbital Modulation of X-rays in CYG X-3}

\author{Shan-Shan Weng\altaffilmark{1, 2}, Shuang-Nan Zhang\altaffilmark{1}, 
Ming-Yu Ge\altaffilmark{1}, Jian Li\altaffilmark{1}, Shu Zhang\altaffilmark{1}}
\altaffiltext{1}{Key Laboratory of Particle Astrophysics, Institute of High 
Energy Physics, Chinese Academy of Sciences, Beijing 100049, China}
\altaffiltext{2}{Department of Physics, Xiangtan University, Xiangtan 411105, 
China}
\email{wengss@ihep.ac.cn; zhangsn@ihep.ac.cn}

\begin{abstract}

We analyze all of the available {\it RXTE} observations of Cyg X-3 in order to 
investigate the connection between the central X-ray source and its surrounding 
environment. The hardness--intensity diagram of Cyg X-3 displays a ``shoe'' 
shape rather than the Q-type shape commonly seen in other black hole X-ray 
binary, and exhibits no apparent hysteresis effect. During the $\gamma$-ray 
outbursts, no existing data are located in the hard and intermediate states, 
which suggest the absence of a significant population of non-thermal electrons 
when the source is in these states. For the first time, we present the orbital 
modulation of the X-ray light curve (LC) of all five states. The different energy 
band LCs are in phase with each other in all five states, and the modulation 
amplitude of both soft and hard X-ray LCs monotonously increases with decreasing 
hardness from hard to soft non-thermal states. We confirm that the modulation 
depth decreases with increasing energy in the hard, intermediate, and very high 
states, as originally reported by Zdziarski et al. However, in the soft non-thermal 
state, the hard X-ray modulation strength significantly increases and is even 
larger than the soft X-ray one. Our results rule out both wind absorption and jet 
origins of the hard X-ray LC modulation in the soft non-thermal state, and
challenge our understanding of the states of Cyg X-3.

\end{abstract}


Online-only material: color figures

\section{Introduction}

More than 40 years ago, the X-ray binary (XRB) Cyg X-3 was first detected by 
Giacconi et al. (1967) during an early rocket flight in 1966; it was then 
observed by all subsequent X-ray missions. It is located in the Galactic plane 
with an estimated distance of $\sim$7-9 kpc (Dickey 1983; Predehl et al. 2000; 
Ling et al. 2009). The infrared (IR) observations indicate a short orbital period 
$P_{\rm orb}=4.8\ {\rm hr}$, and the donor is a Wolf--Rayet (WR) star 
(van Kerkwijk et al. 1996). The dense WR stellar wind prevents the optical photons
from reaching the observer. Therefore, we know practically nothing about its 
system inclination and mass function. It remains uncertain whether the compact 
object is a black hole (BH) or a neutron star (NS, Stark \& Saia 2003; Vilhu 
et al. 2009) up to now. The presence of a BH is favored when the X-ray and radio 
emission and the bolometric luminosity of Cyg X-3 (Szostek \& Zdziarski 2008; 
Zdziarski et al. 2010) are considered.

As one of the brightest XRBs, the X-ray radiation of Cyg X-3 has been studied 
over the past years (e.g. McCollough et al. 1999). Investigating the {\it Rossi 
X-ray Timing Explorer} ({\it RXTE}) All Sky Monitor (ASM) and CGRO/BATSE light 
curves (LCs) of Cyg X-3, Hjalmarsdotter et al. (2008) found that the distribution 
of its ASM count rate is bimodal, with two peaks representing the soft and hard 
states in Cyg X-3, respectively. However, the distribution of ASM hardness and 
BATSE flux did not show the bimodal behavior. Inspecting 42 {\it RXTE}-pointed 
observations, Szostek et al. (2008) further divided the spectra into five groups 
based on their shape, which are ordered by decreasing flux above 20 keV. 
Hjalmarsdotter et al. (2009) interpreted the spectra in terms of Comptonization by 
a hybrid electron distribution and strong Compton reflection. Although the X-ray 
spectra of Cyg X-3 have been intensively  studied, there have not been any
comprehensive studies of the LC modulation in different states as yet.

In addition to the X-ray radiation, luminous radio jets are common features for both
BH and NS XRBs (Coriat et al. 2011). Since first being seen by Gregory et al. (1972), 
Cyg X-3 has frequently been the brightest XRB at radio wavelengths with resolved jets 
(Tudose et al. 2010; Mioduszewski et al. 2001; Miller-Jones et al. 2004), thus 
qualifying the source as a microquasar. The relationship between the radio and X-ray 
was first pointed out by Watanabe et al. (1994), and further studied in the literature 
(e.g., McCollough et al. 1999; Koljonen et al. 2010). By inspecting its radio/X-ray 
correlation, Cyg X-3 can be distinguished from the canonical BH transients, e.g., the 
strong radio flares occur only when Cyg X-3 is in its soft state. Recently, $\gamma$-ray 
flares, which occurred close to the radio flares, were detected in Cyg X-3 by both 
{\it Fermi}/LAT (Abdo et al. 2009) and {\it AGILE} (Tavani et al. 2009).

Being the first microquasar detected in the GeV band, Cyg X-3 draws much attention 
and dense multi-band observations. However, a number of fundamental questions are 
still unanswered. In this work, our aim is to answer the following questions. (1) Does 
the configuration of the stellar wind change during different states? (2) Does the 
hard X-ray LC orbital modulation result from wind absorption? (3) Does the low energy 
tail of the $\gamma$-ray emission contribute to the hard X-ray band? We first analyze 
the archived {\it RXTE} data in the next section. The results and their physical 
implications are presented in Section 3. We also check these results with X-ray 
monitoring data in Section 4. Our summary, conclusion, and discussion are in Section 5.

\section{Data Analysis}

{\it RXTE} has three instruments --- the Proportional Counter Array (PCA), the 
High-Energy X-Ray Timing Experiment, and the ASM. The PCA was designed to cover an 
energy range of 2.0-60.0 keV and consists of five, nearly identical, large-area 
xenon Proportional Counter Units (PCUs). The field of view is about 
$1\degr \times 1\degr$. Each PCU has an area of $\sim$1300 ${\rm cm}^{2}$, for a 
total effective area of 6500 ${\rm cm}^{2}$ (Jahoda et al. 1996). The PCA is now 
well calibrated up to about 50 keV using the latest background models
\footnote{\protect\url{http://heasarc.gsfc.nasa.gov/docs/xte/pca\_bkg\_epoch.html}}
and the response matrices v11.7 released on 2009 August 17 (Shaposhnikov et al. 2012).
\footnote{\protect\url{http://www.universe.nasa.gov/xrays/programs/rxte/pca/doc/rmf/pcarmf-11.7/}}

\begin{figure}
\begin{center}
\includegraphics[scale=0.5]{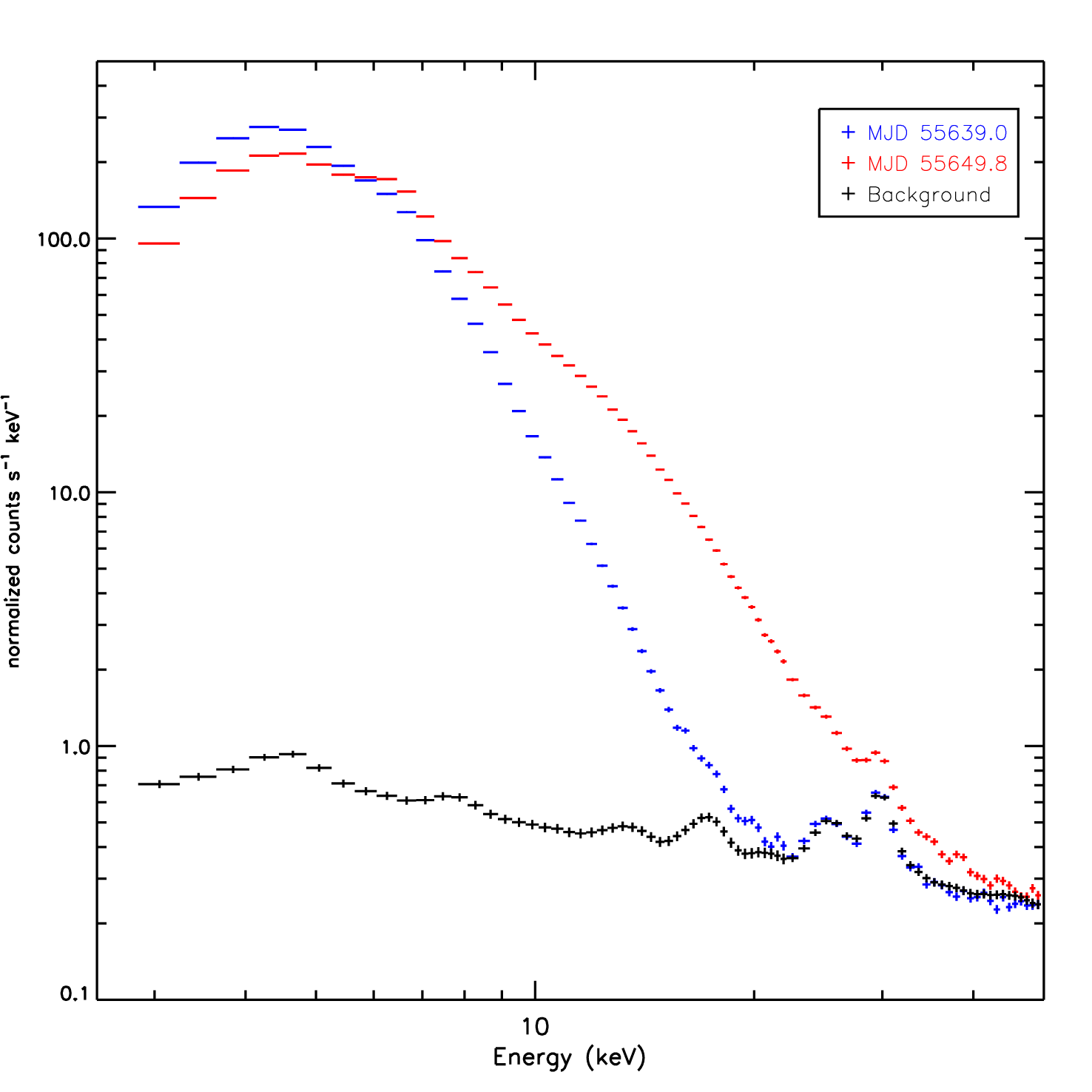}
\caption{\label{fig1} Examples of Cyg X-3 PCA count spectra. The typical source+background
spectrum for the hard state (red symbol) is well above the background spectrum (black symbol).
However, the background spectrum exceeded the source+background spectrum above $\sim$ 20 keV in
the ultrasoft state (blue symbol).}
\end{center}
\end{figure}

Figure 1 is an example of Cyg X-3 PCA observation data, which were used in Corbel 
et al. (2012). The typical source+background count spectra (MJD 55649.8, red symbol) 
are well above the background count spectrum (black symbol). Only when the source 
is faint, as when Cyg X-3 is in the ultrasoft state (MJD 55639, blue symbol; see 
Section 2.2 for the definition of the source states), may the background exceed the 
source+background count spectrum, resulting in a negative count rate at the high 
energy. In our work, we focus on the PCA data and present the energy-dependent 
orbital modulation for different states at energies $\le$ 40 keV.

\subsection{{\it RXTE}/PCA Data Reduction}

The number of {\it RXTE} pointed-observations rapidly increased from 119 to 239 
after 2009 January 1, when the Cyg X-3 $\gamma$-ray emission were confirmed by 
{\it Fermi}/LAT and {\it AGILE} observations. We analyze all {\it RXTE} 
observations of Cyg X-3 with HEASOFT version 6.10.
\footnote{See \protect\url{http://heasarc.gsfc.nasa.gov/docs/software/lheasoft/}} 
For the PCA, we use the Standard2 data from all layers of PCU2, which operated 
during all the observations. The data are filtered with the standard criteria: the 
earth-limb elevation angle is larger than $10\degr$ and the spacecraft pointing 
offset is less than $0.02\degr$. The background files are created using the program 
{\it pcabackest} and the latest bright source background model. Since the data 
used here spanned three {\it RXTE} gain epochs (from epoch 3 to epoch 5) and the 
PCA gain slowly drifted over time, we also analyze the data of Crab to get the 
correct channel-to-energy conversions (Belloni 2010). We first extract the
background-subtracted LCs of Cyg X-3 channel by channel in 16 s time bins, and 
then normalize raw count rates with the quasi-simultaneous observations of Crab. 
Finally, the flux (Crab unit) of the energy band concerned is averaged with the 
inverse squares of the uncertainties as weights over the different channels for 
different gain epochs (Table 1). In this work, the 1$\sigma$ (68\%) errors are given 
with standard error propagation unless otherwise stated.

\begin{deluxetable}{lllllll}
\tabletypesize{\small} \tablewidth{0pt}
\tablecaption{Absolute Channels for Multi-band LCs in Different {\it RXTE} Gain Epochs\label{tab:chanel}}
\tablehead{\colhead{} & \colhead{3-6 keV} &
\colhead{10-15 keV} &\colhead{2-10 keV} & \colhead{10-20 keV} &
\colhead{20-40 keV} &\colhead{15-30 keV}} \startdata

epoch3 & $8-16$ & $27-41$ & $5-27$ & $28-55$ & $56-107$ & $41-79$ \\
epoch4 & $7-14$ & $24-35$ & $5-24$ & $25-47$ & $48-93$ & $35-69$ \\
epoch5 & $7-14$ & $24-36$ & $5-24$ & $25-47$ & $48-95$ & $36-69$ \\

\enddata

\tablecomments{Channels are specified here with respect to the
full 256-channel PCA. The {\it RXTE} epochs are documented at
\url{http://heasarc.nasa.gov/docs/xte/e-c\_table.html}}.

\end{deluxetable}

\subsection{The Hardness--Intensity Diagram}

The hardness--intensity diagram (HID), in which the total count rate is plotted as a
function of hardness, is particularly useful for characterizing the behavior of BH
XRBs (Fender et al. 2004). In order to compare with other BH XRBs, we construct the
HID in a manner similar to Fender et al. (2004) and Koljonen et al. (2010). The
hardness ratio is defined as the ratio of the flux between 3-6 keV and 10-15 keV (the
hardness ratio of Crab is $\sim$ 0.53), and the intensity is calculated as the sum
of the count rates in both bands (Figure 2).

\begin{figure}
\begin{center}
\includegraphics[scale=0.6]{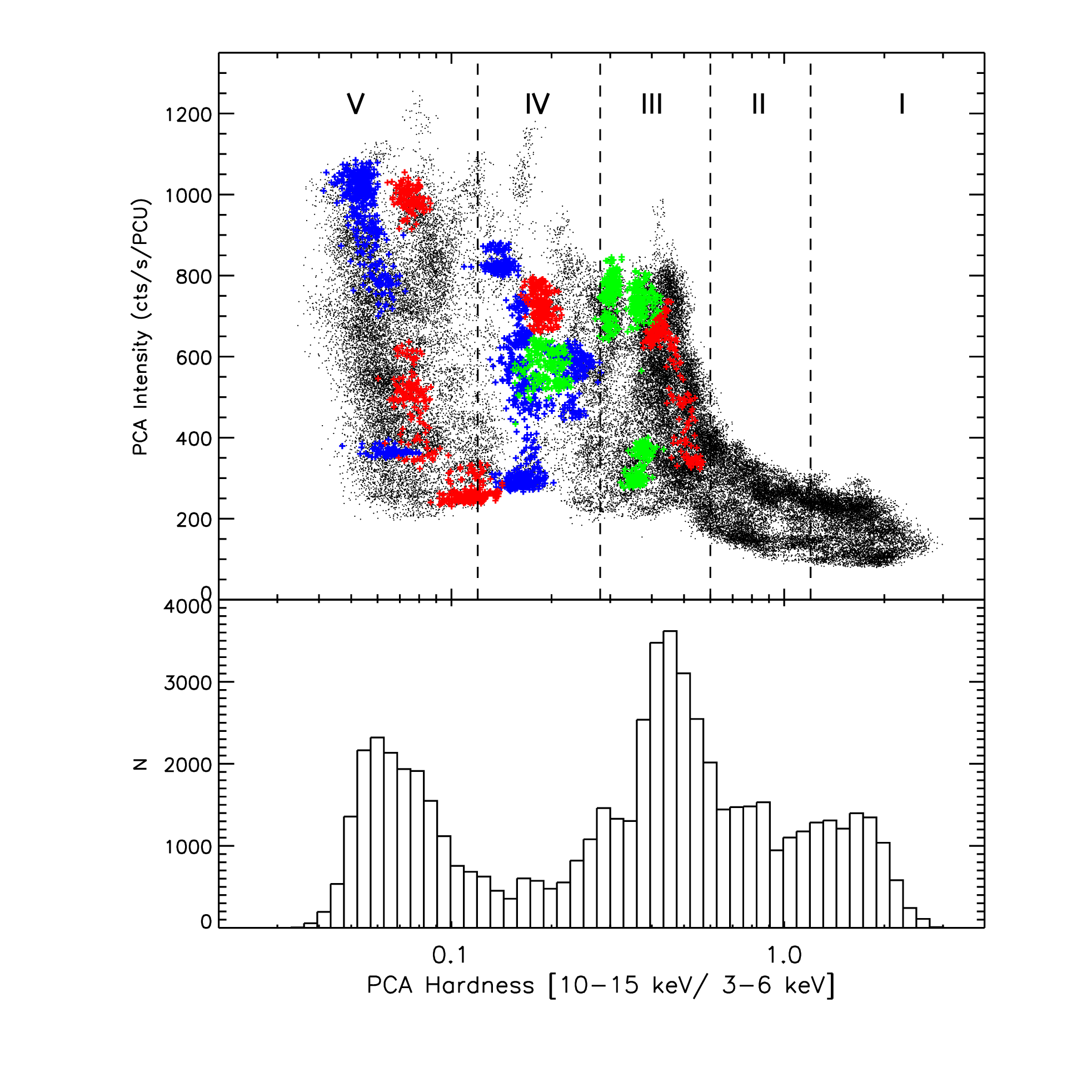}
\caption{\label{fig2} HID (upper panel) and the bimodal hardness distribution
(bottom panel) of Cyg X-3, with bin size of 16 s. The HID is divided into five
regions (I--V), which are occupied by the hard, intermediate, very high, soft
non-thermal, and ultrasoft states, respectively. The 2009, 2010, and 2011 $\gamma$-ray
outbursts data are marked with blue, green, and red crosses, respectively.}
\end{center}
\end{figure}

Following Koljonen et al. (2010), we divide the HID into five regions (I--V), somewhat
arbitrarily, and produce their average spectra. The unweighted average spectra with
data during {\it RXTE}/PCA gain epoch 5, when more than 60\% observations were performed,
are shown in Figure 3. Our results are similar to Szostek et al. (2008), except that no
data above 40 keV are used here. To be consistent with their work, we specify regions
I--V (of Figure 2) as the hard, intermediate, very high, soft non-thermal, and ultrasoft
states (Szostek et al. 2008), respectively.

\begin{figure}
\begin{center}
\includegraphics[scale=0.7]{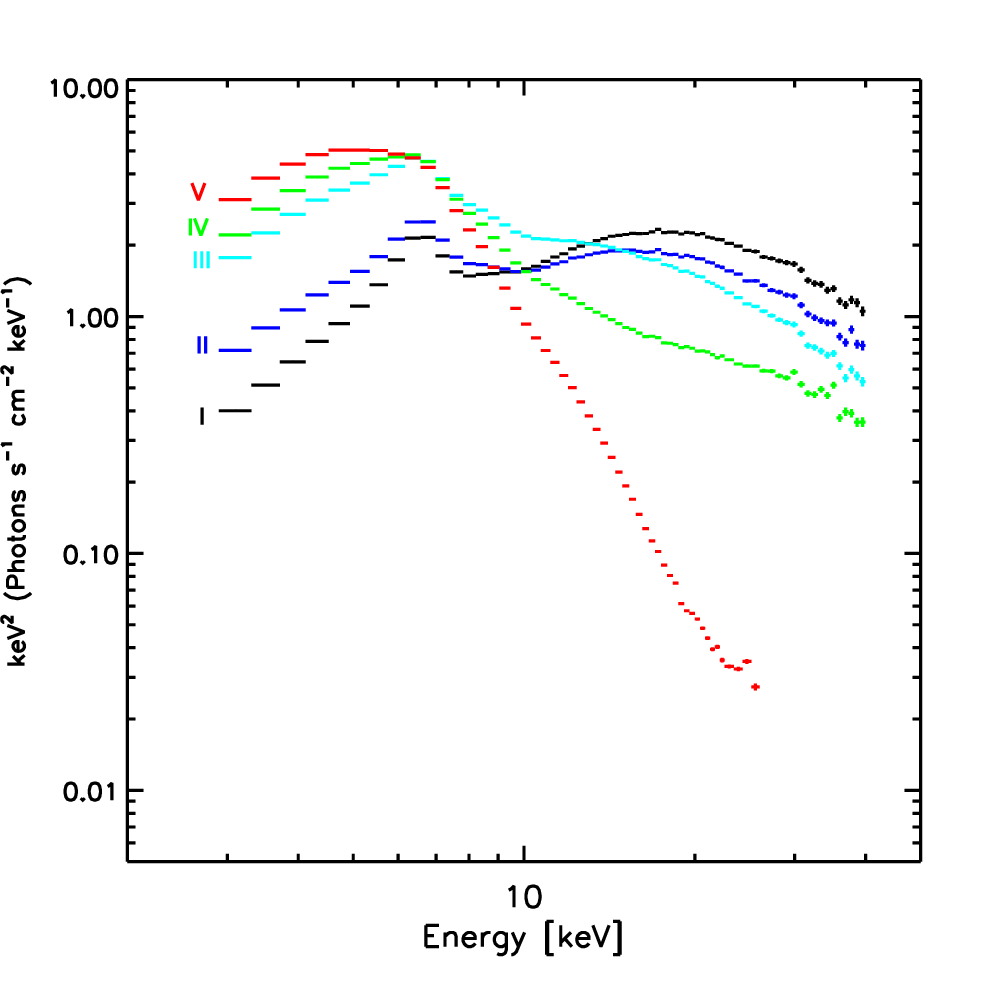}
\caption{\label{fig3} Average spectra of Cyg X-3 in different X-ray states during
the {\it RXTE}/PCA gain epoch 5. This figure is similar to Szostek et al. (2008),
except that no data above 40 keV are used in our work.}
\end{center}
\end{figure}

\subsection{Orbital Modulation in X-Ray LC}

The orbital period $P=4.8\ {\rm hr}$ had been detected in X-ray and IR LCs (Hanson et
al. 2000; Wen et al. 2006; van Kerkwijk 1993), and recently in the GeV band (Abdo et al.
2009). Wen et al. (2006) derived a period of $P = 0.1996907(7) {\rm days} $ from 8.5 yr
of {\it RXTE}/ASM data. In addition, the evolution of the orbit had been studied in
literature (Elsner et al. 1980; van der Klis \& Bonnet-Bidaud 1989). The large value
of the time derivative of the period ($\dot{P}$, $\dot{P}/P \simeq 1\times 10^{-6} {\rm yr}^{-1}$)
has been suggested to be linked to the binary evolution of the system (Singh et al.
2002). Because the data used here span more than 5300 days, the derivative $\dot{P}$
can significantly affect the orbital modulation (Figure 4). Therefore, a quadratic
ephemeris is adopted for phase determination in our work.

\begin{figure}
\begin{center}
\includegraphics[scale=0.6]{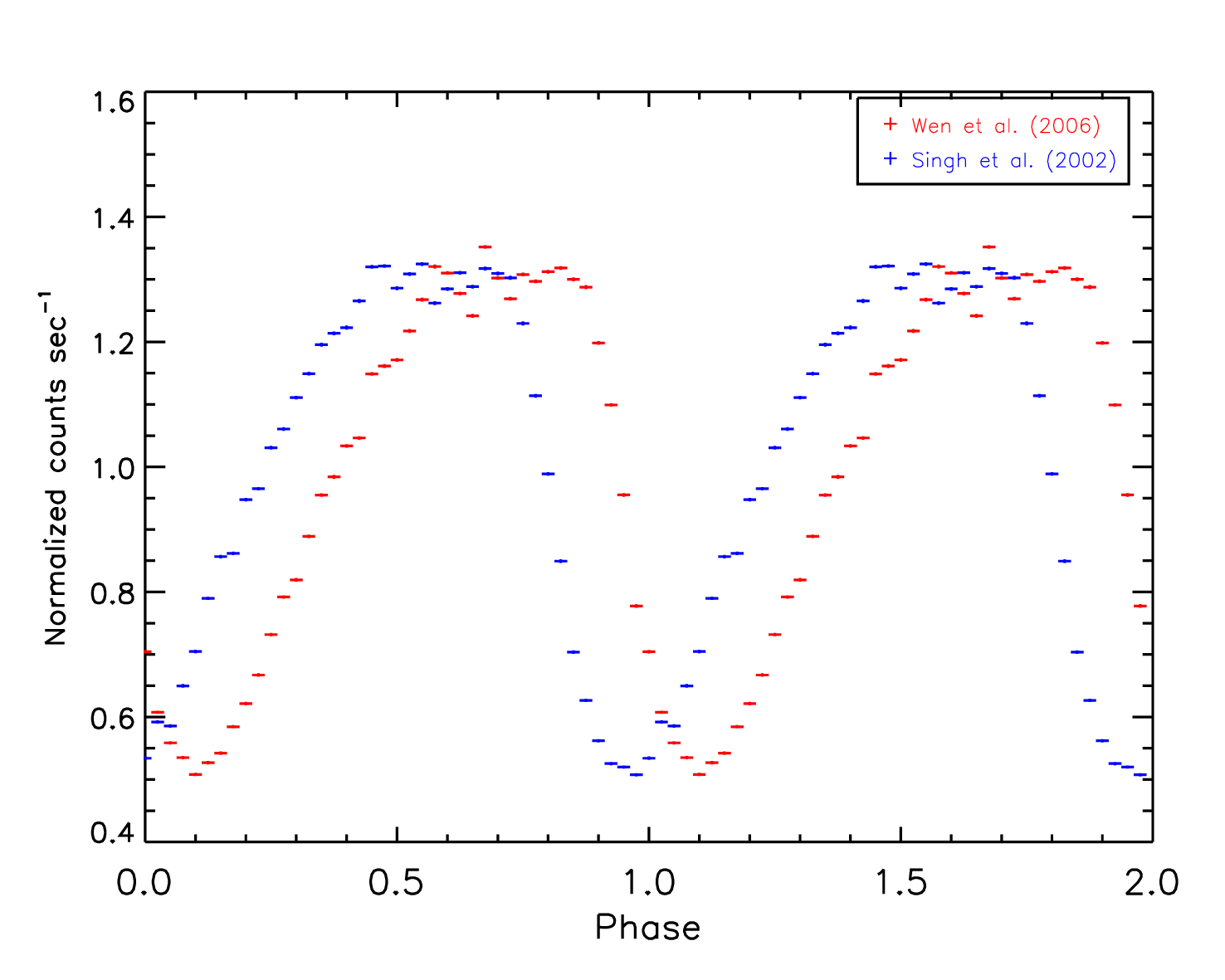}
\caption{\label{fig4} ASM dwell data over the entire lifetime of {\it RXTE} are folded over
the orbital period using the quadratic ephemeris (blue symbol; Singh et al. 2002) and the
linear ephemeris (red symbol; $T_0=40949.392 [{\rm MJD}$], $P = 0.19969077 {\rm d} $; Wen
et al. 2006), respectively.}
\end{center}
\end{figure}

We first calculate the phases of the data points using the ephemeris of Singh et al. (2002),
\begin{equation}
T_0=40949.392\, [{\rm MJD}], \, P_0=0.19968443\,{\rm d},\, \dot{P} =5.76\times 10^{-10},
\end{equation}
where $P_0$ is the period at the reference time $T_0$. The source is variable, and
the weighted average modulation profiles are biased by the higher weight data points.
Alternatively, the folded LCs are given by the unweighted averages of the fluxes in each phase
bin. In this work, we take two approaches to define the modulation amplitude, which are
basically equivalent.

The first approach fits the folded LCs with the trigonometric function
\begin{equation}
 f(\phi)= A*{\rm sin}[(\phi-\phi_0)*2\pi-\pi/2]+C,
\end{equation}
where the orbital phase $\phi$ is within the 0-1 interval. There are three free parameters
in the function---$C$ fluctuates near the mean value $\sim 1$, $\phi_0$ corresponds to the
minimum flux, and $A$ represents the modulation amplitude. However, because the LC
profiles are very complex, the single sinusoid cannot accurately describe them, and may
miss the actual flux extrema.

The second approach is that the modulation amplitude is calculated as
$F = (f_{\rm max}-f_{\rm min})/(f_{\rm max}+f_{\rm min})$, where $f_{\rm max}$ and
$f_{\rm min}$ are the maximum and minimum fluxes. $F$ is independent of the model, but might
overestimate the modulation amplitude due to fluctuations of the maximum (and/or minimum)
fluxes.

\begin{figure}
\begin{center}
\includegraphics[scale=0.5]{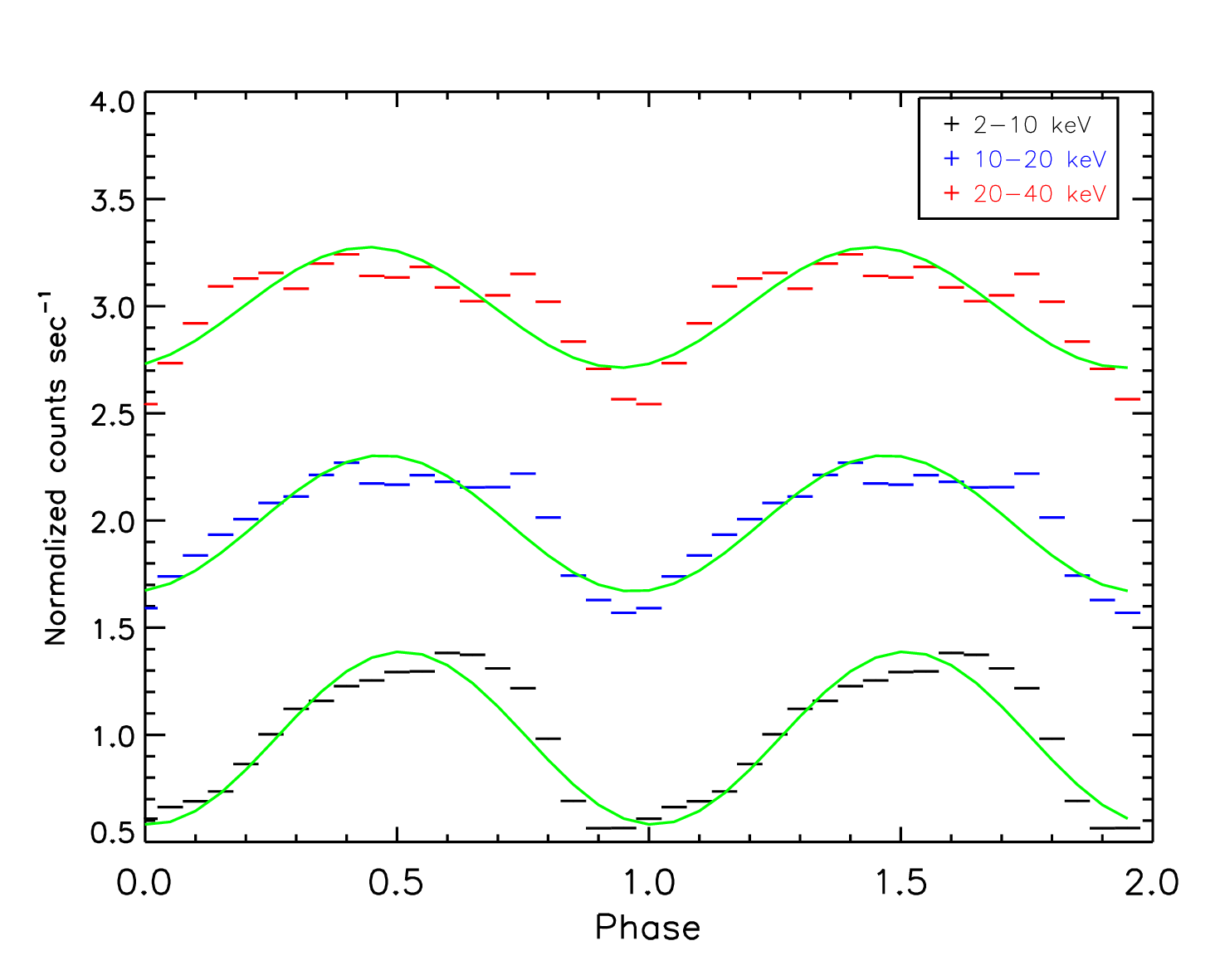}
\caption{\label{fig5} PCA LCs of 2-10 keV (black points), 10-20 keV (blue points),
and 20-40 keV (red points) folded over its orbital period, respectively. The 10-20 keV
and 20-40 keV LCs are shifted up by 1 and 2, respectively, for clarity. The LCs are
also fitted with the trigonometric function (green lines).}
\end{center}
\end{figure}

Because of the spectral pivoting behavior of Cyg X-3 at 10-20 keV in different states, we
create the PCA 2-10 keV, 10-20 keV, and 20-40 keV LCs to investigate the
energy-dependent orbital modulation. Figure 5 shows that the modulation amplitude
decreases with increasing energy, $A_{2-10 \rm keV} = 0.44\pm0.04$, $A_{10-20 \rm keV} = 0.23\pm0.03$
and $A_{20-40 \rm keV} = 0.15\pm0.05$ ($F_{2-10 \rm keV} = 0.43\pm0.01$,
$F_{10-20 \rm keV} = 0.31\pm0.01$ and $F_{20-40 \rm keV} = 0.33\pm0.01$) for 2-10 keV,
10-20 keV and 20-40 keV, respectively. However, the strong fluctuations exhibited in
the 20-40 keV LC indicate that the modulation varies with the states. Because of the
complicated spectral states in Cyg X-3, the physical meaning of orbital modulation for
all PCA data is unclear.

\subsection{Energy-dependent Orbital Modulation for Different States}

Because the orbital period of Cyg X-3 is unusually short, the accumulated exposure time
of the {\it RXTE}-pointed data of $\sim$ 980 ks covers more than 56 times its
orbital period. Thus, it allows us to study the soft and hard X-ray orbital modulation
in different states with a coverage time of at least $\sim$ 5.4 times the orbital period
(region IV of Figure 2) per state. The 2-10 keV, 10-20 keV, and 20-40 keV LCs are created
for all five states (Figure 6), and their modulation parameters are listed in Table 2.
However, due to the background being over subtracted (Figure 1), the count rate of the
20-40 keV band becomes negative in the ultrasoft state, and the modulation parameters
are unavailable.

\begin{figure}
\begin{center}
\includegraphics[scale=0.7]{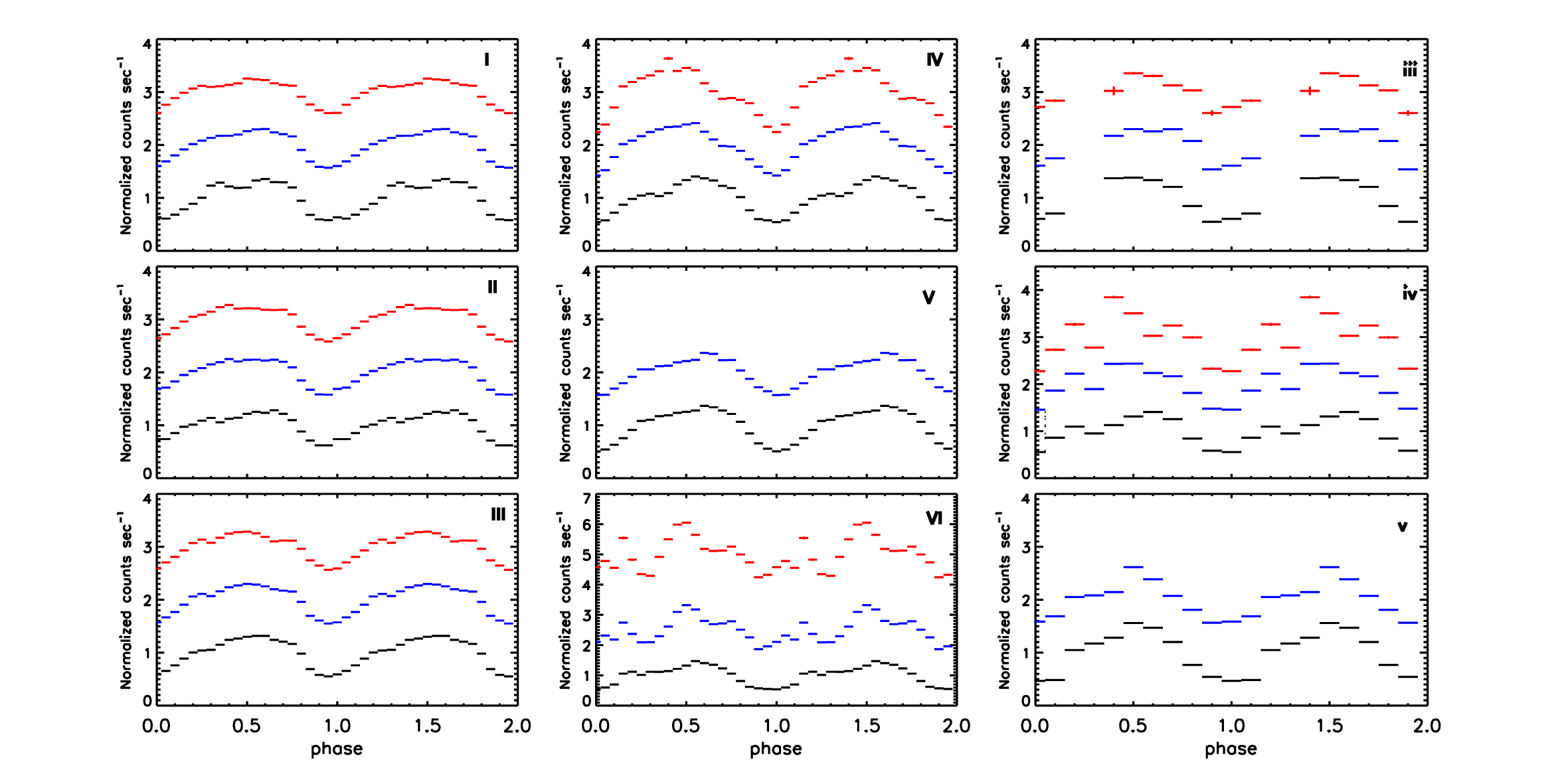}
\caption{\label{fig6} Panels I--V: the LCs of 2-10 keV (black points), 10-20 keV (blue
points), and 20-40 keV (red points) folded over its orbital period for regions I--V of
Figure 2, respectively. The 10-20 keV and 20-40 keV LCs in panels I, II, III, IV,
V, VI, iii, iv, and v are shifted up by (1, 1, 1, 1, 1, 1.5, 1, 1, 1) and
(2, 2, 2, 2, --, 4, 2, 2, --), respectively, for clarity. Panel VI: the LCs of 2-10 keV,
10-20 keV, and 20-40 keV during the 2009, 2010, and 2011 $\gamma$-ray active periods.
During $\gamma$-ray outbursts, all existing data are located in the very high, soft
non-thermal, and ultrasoft states, and their LCs are plotted in panels iii--v,
respectively.}
\end{center}
\end{figure}

\begin{deluxetable}{cccccc}
\tabletypesize{\footnotesize} \tablewidth{0pt} \tablecaption{Orbital Modulation for Different States}
\tablehead{\colhead{} & \colhead{Region I} & \colhead{Region II} &\colhead{Region III} & \colhead{Region IV} &
\colhead{Region V} \\ \colhead{} & \colhead{Hard State} & \colhead{Intermediate State} &\colhead{Very High
State} & \colhead{Soft Non-thermal State} & \colhead{Ultrasoft State}} \startdata

 & &  &  2-10 keV       & &   \\

$A$ & $0.39\pm0.03$ & $0.30\pm0.03$ & $0.38\pm0.03$ & $0.41\pm0.03$ & $0.39\pm0.03$  \\
$\phi_{0}$ & $0.00\pm0.01$ & $0.96\pm0.02$ & $0.00\pm0.01$ & $0.01\pm0.02$ & $0.02\pm0.01$  \\
$F$ & $0.40\pm0.01$ & $0.35\pm0.01$ & $0.41\pm0.01$ & $0.44\pm0.01$ & $0.45\pm0.01$  \\

\hline
 & &  &  10-20 keV      & &   \\

$A$ & $0.36\pm0.03$ & $0.34\pm0.03$ & $0.37\pm0.03$ & $0.48\pm0.04$ & $0.34\pm0.03$  \\
$\phi_{0}$  & $0.97\pm0.01$ & $0.95\pm0.02$ & $0.99\pm0.01$ & $0.96\pm0.01$ & $0.03\pm0.01$  \\
$F$ & $0.39\pm0.01$ & $0.37\pm0.01$ & $0.40\pm0.01$ & $0.54\pm0.01$ & $0.40\pm0.01$  \\

\hline
 & &  &  20-40 keV       & &   \\

$A$ & $0.30\pm0.03$ & $0.32\pm0.03$ & $0.34\pm0.03$ & $0.57\pm0.06$ & ...  \\
$\phi_{0}$  & $0.95\pm0.01$ & $0.95\pm0.01$ & $0.98\pm0.01$ & $0.95\pm0.02$ & ...  \\
$F$ & $0.35\pm0.01$ & $0.37\pm0.01$ & $0.39\pm0.01$ & $0.74\pm0.03$ & ...  \\

\hline
 & &  &  15-30 keV      & &   \\

$A$ & $0.32\pm0.04$ & $0.33\pm0.03$ & $0.35\pm0.03$ & $0.52\pm0.04$ & ...  \\
$\phi_{0}$  & $0.96\pm0.01$ & $0.95\pm0.01$ & $0.98\pm0.01$ & $0.95\pm0.02$ & ...  \\
$F$ & $0.36\pm0.01$ & $0.37\pm0.01$ & $0.39\pm0.01$ & $0.64\pm0.01$ & ...  \\

\enddata

\tablecomments{The fitted modulation parameters and their 1$\sigma$ errors. Due to
the background being over-subtracted, we cannot measure the modulation parameters for
20-40 keV in the ultrasoft state.}

\end{deluxetable}

Before obtaining the final results, we carry out a series of tests.

\begin{itemize}
  \item Boundaries between the states:
  the boundaries between the states are somewhat arbitrary. After shifting the
  boundaries by a factor of 10\%, the results are approximately the same.

  \item Energy band of the hard X-ray:
  because of the background issue at 20-40 keV, the lower energy band (15-30
  keV) LCs are created and used to study the hard X-ray LC modulation.

  \item Bin size of LCs:
  the short timescale variability might affect the modulation profiles and
  amplitude. The LCs are rebinned to 128 s (or 256 s) and then folded.

\end{itemize}

We find that all these elements and their combinations modify the modulation
amplitudes only within 25\%, but the modulation evolution patterns (see Section 3)
of the soft and hard X-ray LCs are essentially unchanged. We fold the 256 s
time-bin-size LCs and present the results in Table 3.

Since its first detection in the GeV band in 2009, {\it Fermi}/LAT and {\it AGILE} 
have detected several activities from Cyg X-3, which were partly covered by 
{\it RXTE} data from 2009 June 14 to 2009 July 27 (Abdo et al. 2009), 2010 May 29 
to 2010 June 3 (Williams et al. 2011), and 2011 March 20 to 2011 March 24 
(Corbel et al. 2012) with a total exposure time $\sim$ 47.7 ks (2.76 orbits). The 
data during the 2009, 2010, and 2011 $\gamma$-ray active periods are marked by 
blue, green, and red crosses in the HID, respectively. The modulation profile of 
the LCs in 10-20 keV and 20-40 keV during these epochs deviates from the sinusoidal 
trace (panel VI of Figure 6). Note that these $\gamma$-ray outbursts took place 
during different radio/X-ray states, and the accumulated exposure time 
($\sim$ 47.7 ks) is relatively short. In this case, for the given phases, the LCs 
may be dominated by the data of different states. Thus, we further create the LCs 
for the very high, soft non-thermal, and ultrasoft states (see panels iii--v of 
Figure 6), respectively.

\begin{deluxetable}{cccccc}
\tabletypesize{\footnotesize} \tablewidth{0pt} \tablecaption{Modulation Parameters 
for Different States \label{tab:modulation}} 
\tablehead{\colhead{} & \colhead{Region I} & \colhead{Region II} &\colhead{Region III} 
& \colhead{Region IV} & \colhead{Region V} \\ \colhead{} & \colhead{Hard State} & 
\colhead{Intermediate State} &\colhead{Very High State} & \colhead{Soft Non-thermal 
State} & \colhead{Ultrasoft State}} \startdata

 & &  &  2-10 keV       & &   \\

$A$ & $0.38\pm0.04$ & $0.30\pm0.03$ & $0.40\pm0.03$ & $0.43\pm0.03$ & $0.39\pm0.03$  \\
$\phi_{0}$ & $0.00\pm0.01$ & $0.96\pm0.02$ & $0.00\pm0.01$ & $0.00\pm0.01$ & $0.03\pm0.01$  \\
$F$ & $0.37\pm0.01$ & $0.40\pm0.01$ & $0.40\pm0.01$ & $0.46\pm0.01$ & $0.45\pm0.01$  \\

\hline
 & &  &  10-20 keV      & &   \\

$A$ & $0.35\pm0.03$ & $0.34\pm0.03$ & $0.38\pm0.03$ & $0.50\pm0.04$ & $0.30\pm0.03$  \\
$\phi_{0}$  & $0.97\pm0.01$ & $0.95\pm0.02$ & $0.99\pm0.01$ & $0.97\pm0.01$ & $0.03\pm0.01$  \\
$F$ & $0.38\pm0.01$ & $0.38\pm0.01$ & $0.39\pm0.01$ & $0.57\pm0.01$ & $0.36\pm0.01$  \\

\hline
 & &  &  20-40 keV      & &   \\

$A$ & $0.28\pm0.03$ & $0.30\pm0.02$ & $0.31\pm0.02$ & $0.56\pm0.06$ & ...  \\
$\phi_{0}$  & $0.95\pm0.01$ & $0.94\pm0.01$ & $0.98\pm0.01$ & $0.97\pm0.01$ & ...  \\
$F$ & $0.33\pm0.01$ & $0.34\pm0.01$ & $0.34\pm0.01$ & $0.69\pm0.04$ & ...  \\

\hline
 & &  &  15-30 keV     & &   \\

$A$ & $0.31\pm0.04$ & $0.32\pm0.02$ & $0.35\pm0.03$ & $0.53\pm0.05$ & ...  \\
$\phi_{0}$  & $0.96\pm0.01$ & $0.95\pm0.01$ & $0.98\pm0.01$ & $0.97\pm0.01$ & ...  \\
$F$ & $0.35\pm0.01$ & $0.35\pm0.01$ & $0.37\pm0.01$ & $0.63\pm0.01$ & ...  \\

\enddata

\tablecomments{The same modulation parameters for different states as in Table 2, but now 
the LCs are rebinned to 256 s.}

\end{deluxetable}

\section{Results and Physical Implications}

\subsection{No Hysteresis in Cyg X-3}

The HID of Cyg X-3 displays a ``shoe'' shape rather than the Q-type shape commonly seen
in other BH XRBs (Chen et al. 2010), and exhibits no apparent hysteresis effect, which is
frequently seen in other BH XRBs (Yu \& Yan 2009) and a few NS XRBs (see e.g., Figure 1 in
Weng \& Zhang 2011). In the bottom panel of Figure 2, the PCA hardness displays a bimodal
distribution, which is similar to the case in the ASM hardness of Cyg X-1 (Hjalmarsdotter
et al. 2008). This is possibly due to its spectral pivot energy of $\sim$ 10-15 keV, which
is beyond the ASM band. However, a caveat is that the PCA observations were not performed
uniformly, and that the bimodal distribution might be artificial.

The strong stellar wind in Cyg X-3 provides abundant accretion material that makes the
source persistent, i.e., the source never enters into the usual quiescent state. As a
result, the hard state is restricted in the limited flux range. The hysteresis is thought
to be triggered by the dramatic changes in accretion flow during the hydrogen ionization
instability. Cyg X-3 should have a small disk due to the tight orbit and the wind
accretion, and the accretion rate is higher than the critical accretion rate, making the
disk temperature above the hydrogen ionization temperature everywhere (Done et al. 2007);
this is probably why it does not show hysteresis. Another plausible potential candidate
mechanism for suppressing the hysteresis is that the strong stellar wind collides with the
accretion disk, which heats up the outer disk.

\subsection{The X-Ray Behavior during the $\gamma$-Ray Outbursts}

Whether the low energy tail of the $\gamma$-ray emission can give a contribution to the
hard X-ray band is still under debate (Zdziarski et al. 2012b). During the $\gamma$-ray
outbursts, all existing data are located in the very high, soft non-thermal, and ultrasoft
states (regions III, VI, and V of Figure 2), but none in the hard and intermediate states
(regions I, and II of Figure 2). This suggests the absence of a significant population of
non-thermal electrons in the system when the source is in the hard and intermediate states
(Hjalmarsdotter et al. 2009; Acciari et al. 2011). When entering into the softer states,
i.e., the very high, soft non-thermal, and ultrasoft states, the source may launch a strong
radio flare. The relativistic electrons in the jets can up-scatter the dense UV photons
emitted by the WR star, and produce the detected GeV emission (Dubus et al. 2010).
Figure 6 shows that the modulations for the data simultaneous and not simultaneous with
observations of $\gamma$-ray outbursts give the same profiles. This indicates that the X-ray
and $\gamma$-ray emissions are created in different regions.

\subsection{Soft and Hard X-Ray LCs Modulation}

First, we construct both the soft and hard X-ray modulation profiles for all five states
using all PCA data. The modulation amplitude monotonously increases from the hard to soft
non-thermal states (regions I -- IV). The hard X-ray has a smaller modulation depth than
the soft X-ray one in the hard, intermediate, and very high states (regions I -- III), 
which are consistent with those presented in Zdziarski et al. (2012a). However, the hard 
X-ray modulation strength increases significantly, and the relation between modulation 
amplitude and energy reverses in the soft non-thermal state (region IV).

If X-ray LC modulation is the result of the absorption in the stellar wind, the 
asymmetric profile, with a slow rise and a fast decline, implies a complicated wind 
structure. When the stellar wind becomes stronger, it provides both a denser absorbing 
medium and larger accretion rate, and therefore stronger modulation in softer state. At 
various photon energies, a given state is governed by the absorption cross sections. Both 
the photoionization and Compton scattering have cross sections decreasing with increasing 
X-ray energy (Zdziarski et al. 2012a). The hard X-ray LC shows stronger modulation than 
the soft X-ray LC, and therefore points to another origin of orbital modulation in the 
soft non-thermal state.

Both IR and $\gamma$-ray emissions were found to be also strongly modulated on the orbital
period (Fender et al. 1999; Abdo et al. 2009). The IR emission was thought to arise from the
WR stellar wind, which cannot emit X-ray emissions. On the other hand, the stellar radiation
field is not isotropic for the relativistic electrons in the jet, and results in a
modulation of the $\gamma$-ray emission (Dubus et al. 2010). In this scenario, the jet
modulation profiles have opposed phases with respect to those in soft X-ray LCs, and have the
maximum around the superior conjunction. However, no significant phase shift between the soft
and hard X-ray LCs is detected in our data (Table 3 and Figure 6). Moreover, the modulation
for the data simultaneous with the observations of $\gamma$-ray outbursts has the same
features as the modulation for the other data, and thus the hard X-ray ($\leq$ 40 keV) cannot 
be the low energy tail of the $\gamma$-ray emission.

\section{State-Dependent Orbital Modulation from X-ray Monitoring data}

The public X-ray monitoring data from the {\it Swift} Burst Alert Telescope (BAT)
\footnote{\protect\url{http://swift.gsfc.nasa.gov/docs/swift/results/transients}} and dwell 
data from {\it RXTE}/ASM \footnote{\protect\url{http://xte.mit.edu/ASM\_lc.html}} are also 
used to study the orbital modulation in different states. Cyg X-3 can transit from one state 
to another within one to two days, and different states may provide the same count rate, 
though a softer state generally emits more soft X-ray and less hard X-ray count rates (bottom 
panel of Figure 7; see also Figure 4 of Corbel et al. 2012). However, both BAT and ASM data 
only give the count rates in a single channel and lack spectral information, thus forbidding 
us from looking into the spectra in detail. We therefore cannot judge the source state without 
the help of the PCA data. In this case, we only use the data that are simultaneous with the 
PCA pointed observations, and assume that the source stays at the state given by the PCA data 
for $\pm 1$ day (Figure 7). With these data selection criteria, the remaining data are 
relatively sparse and the modulation profiles might be governed by the aperiodic variability. 
Figure 8 shows the ASM and BAT LCs folded over its orbital period for all five states; the 
modulation parameters are listed in Table 4. We find that the ASM LCs have a level of 
modulation similar to the PCA 2--10 keV LCs, but the modulation amplitudes of BAT LCs are 
systematically lower than the PCA 15-30 keV LCs. Note that the BAT data encounter the same 
background issue as the PCA data in the ultrasoft state---the count rate becomes negative (red
points in the middle panel of Figure 7), and thus no modulation parameter is available.

\begin{figure}
\begin{center}
\includegraphics[scale=0.5]{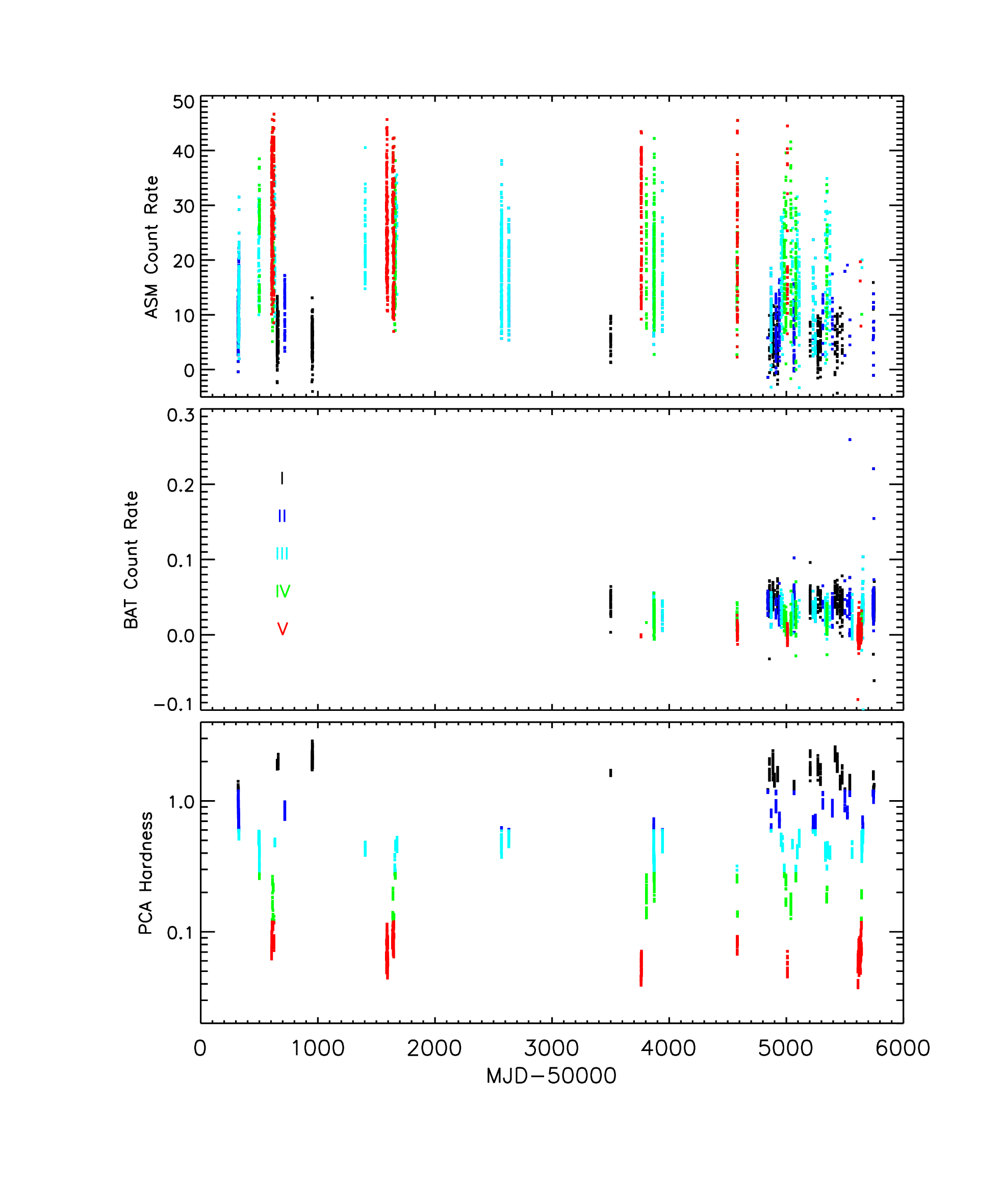}
\caption{\label{fig7} From top to bottom panels are shown the ASM (1.5-12 keV) and BAT (15-50 keV)
light curves, and the temporal evolution (256 s time bin size) of the PCA hardness ratios
(Figure 2), respectively. Note that we only use the ASM and BAT data which are simultaneous with
the PCA-pointed observations. The data of states I, II, III, IV, and V are marked with red,
green, cyan, blue and black symbols (see the text for description of the states), respectively.}
\end{center}
\end{figure}

To check the extrema in the soft non-thermal state, we perform two more tests:

1. Because of poor statistics, the LCs are folded with a small phase bin number ($k = 10$) 
per period (Figure 8). Because using too large phase bin may suppress the orbital modulation, 
we fold the ASM and BAT LCs with $k = 15, 20,$ and 30, and obtain the modulation amplitudes 
$F_{\rm ASM} = 0.49\pm0.01, 0.51 \pm0.01, 0.55\pm0.02,$ and
$F_{\rm BAT} = 0.60\pm0.06, 0.63\pm0.07, 0.67 \pm0.08$ in the soft non-thermal state, 
respectively.

2. Assuming that the source stays at the state given by the PCA data for $\pm 0.5$ and $\pm 2.0$ 
days, the modulation amplitudes for the ASM and BAT data are $F_{\rm ASM} = 0.48\pm0.02, 
0.47\pm0.01,$ and $F_{\rm BAT} = 0.56\pm0.07, 0.52\pm0.04$ ($k =15$ adopted here), respectively.

\begin{figure}
\begin{center}
\includegraphics[scale=0.6]{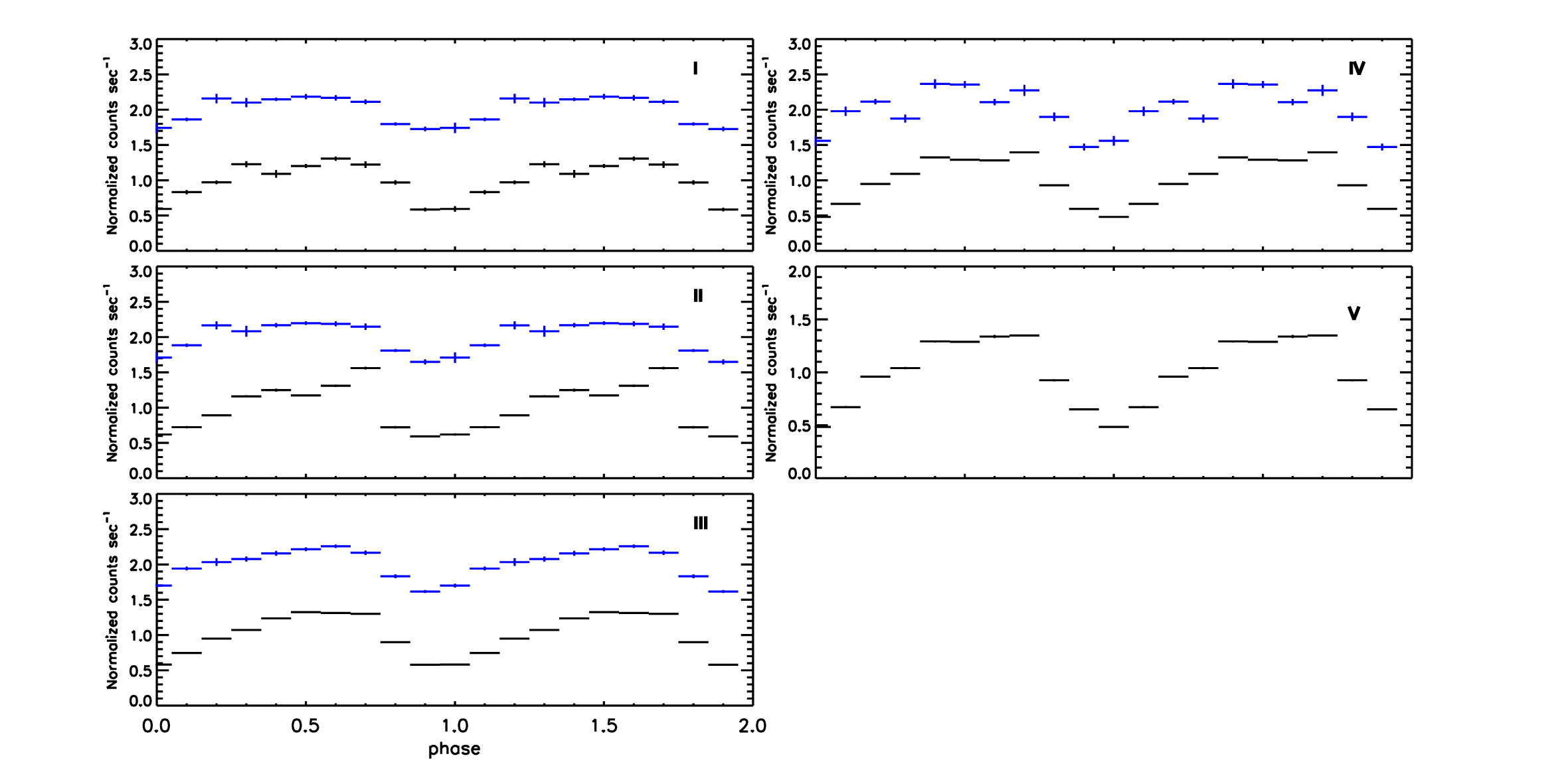}
\caption{\label{fig8} Panels I--V: The LCs of ASM (black points) and BAT (blue points) folded over
its orbital period for regions I--V of Figure 2, respectively. The BAT LCs are shifted up by 1
for clarity.}
\end{center}
\end{figure}

Despite systematic bias and higher statistical fluctuations, the monitoring data exhibit an
evolutionary pattern similar to PCA data during these states. From the hard to soft non-thermal 
states, the modulation amplitude of ASM data  $F_{\rm ASM}$ increases from $0.38\pm0.02$ to 
$0.49 \pm0.01$, whereas $F_{\rm BAT}$ increases much more, from $0.24\pm0.03$ to $0.49\pm0.05$. 
The BAT LC exhibits higher (or at least the same) orbital modulation amplitude than the ASM LC 
in the soft non-thermal state. This implies that orbital modulation of the hard X-ray emission 
in the soft non-thermal state has a different origin from wind absorption.

\begin{deluxetable}{cccccc}
\tabletypesize{\footnotesize} \tablewidth{0pt} \tablecaption{Orbital Modulation in ASM and 
BAT Data} \tablehead{\colhead{} & \colhead{Region I} & \colhead{Region II} &\colhead{Region III} 
& \colhead{Region IV} & \colhead{Region V} \\ \colhead{} & \colhead{Hard State} & 
\colhead{Intermediate State} &\colhead{Very High State} & \colhead{Soft Non-thermal State} & 
\colhead{Ultrasoft State}} \startdata

 & &  &  {\it RXTE}/ASM 1.5-12 keV       & &   \\

$A$ & $0.32\pm0.06$ & $0.38\pm0.09$ & $0.38\pm0.05$ & $0.43\pm0.06$ & $0.42\pm0.06$  \\
$\phi_{0}$ & $0.98\pm0.03$ & $0.00\pm0.03$ & $0.00\pm0.02$ & $0.01\pm0.02$ & $0.02\pm0.03$  \\
$F$ & $0.38\pm0.02$ & $0.45\pm0.01$ & $0.39\pm0.01$ & $0.49\pm0.01$ & $0.47\pm0.01$  \\

\hline
 & &  &  {\it Swift}/BAT 15-50 keV      & &   \\

$A$ & $0.26\pm0.04$ & $0.28\pm0.06$ & $0.27\pm0.05$ & $0.37\pm0.10$ & ...  \\
$\phi_{0}$  & $0.92\pm0.03$ & $0.93\pm0.03$ & $0.96\pm0.04$ & $0.95\pm0.04$ & ...  \\
$F$ & $0.24\pm0.03$ & $0.30\pm0.02$ & $0.34\pm0.02$ & $0.49\pm0.05$ & ...  \\

\enddata

\tablecomments{Due to the background being over-subtracted, we cannot measure the modulation
parameters for BAT LC in the ultrasoft state.}

\end{deluxetable}

\section{Summary, Conclusion and Discussion}

In this paper, we analyzed the orbital modulations of both the soft and hard X-ray LCs of Cyg X-3,
and their relation with other observation data, including the GeV observations. Our main
results in this work are summarized as follows:

1. The HID of Cyg X-3 displays a ``shoe'' shape rather than the Q-type shape commonly seen
in other BH XRBs, and exhibits no apparent hysteresis effect, which is consistent with
previous work (Hjalmarsdotter et al. 2008; Koljonen et al. 2010).

2. Owing to a lack of a significant population of non-thermal electrons in the system, there is no
$\gamma$-ray outburst detected in the hard and intermediate states.

3. We confirm that both the soft and hard LCs have stronger orbital modulation in softer states,
and that the modulation amplitude decreases with increasing energy in the hard, intermediate, and
very high states; this was first reported by Zdziarski et al. (2012a). However, in the soft
non-thermal state, the modulation amplitude of the hard X-ray LC becomes larger than that of the soft
X-ray, which is in contradiction with the wind absorption scenario.

There is no clear evidence of jet contribution to the observed X-ray radiation, since the
modulation for the data that are simultaneous with the observations of $\gamma$-ray outbursts has the
same features as the modulation for the other data. The result indicates that the X-ray and
$\gamma$-ray emissions are created in different regions. Because the hard X-ray LCs are always in
phase with the soft X-ray LCs, but out of phase with $\gamma$-ray emission, we can further rule out
the jet origin of hard X-ray and its orbital modulation.

Our results of state-dependent orbital modulation present a challenge for us in understanding
the state of Cyg X-3.

\acknowledgments{} We thank the anonymous referee for constructive criticism and
suggestions, that have allowed us to significantly improve the presentation of this paper.
This research has made use of data obtained from the High Energy Astrophysics Science Archive
Research Center (HEASARC), provided by NASA's Goddard Space Flight Center. S.S.W. thanks
Jinlu Qu and Wei Cui for many useful suggestions. We acknowledge partial funding support by
the 973 Program of China under grant 2009CB824800, the National Natural Science Foundation
of China under grant Nos. 11133002, 10725313, 11073021, 11103020, 11173024, the Qianren
start-up grant 292012312D1117210, the grants AYA2012-39303, SGR2009-811, and iLINK2011-0303.

\end{document}